\documentstyle{article}

\begin{document} 
\title{EFFECT OF DENSITY CORRELATIONS ON THE COHERENCY 
OF RELATIVISTIC BUNCH RADIATION}
\author{R. V. Tumanian \thanks{ raphael@star.yerphi.am}, L. A. Gevorgian ,
Yerevan Physics Institute,Armenia,\\ Report at NATO Workshop Nor-Hamberd, 
Yerevan, Armenia, 25-29 June, Ed. by H. Wiedemann, 'Electron-Photon 
Interaction in the Dense Media', p.p. 295-301}

\maketitle
\begin{abstract}

The coherent radiation of the electron beam with account of particle 
position correlations because of strong Coulomb interaction is 
considered. For the first time it is shown that the coherency of the  
radiation of the bunch with account of correlations  depend on the 
mean inter particle distance (mean density of the bunch). The conditions 
of coherency are found. The influence of deviations from mean distance 
on the radiation coherency is considered.

\end{abstract}

\section{Introduction}
 In recent years very high density or small size bunches become to possible 
to obtain \cite{1,2}. It is clear that in such dense bunches  the self space charge 
forces of the bunch particles (electrons) play significant influence on the 
bunch parameters and particles motion. However, in this paper we investigate 
the useful effects of the space charge instead of  deleterious effects of that.
  In the last two decades very important  experimental \cite{9,10} and 
theoretical \cite{11-14} results have been achieved in investigations of 
the ordered or crystalline beams. The possibility of such new and interesting 
state of matter is become real in the storage rings of charged particles
and as we show below in linear accelerators with high density beams too.
The case of real storage ring lattice was considered in \cite{15}, where
it was shown that strong cooling is necessary for attaining crystalline 
beams in the storage rings. In difference from storage rings in the 
\cite{18}it was shown that energy spread,sizes and particles motion 
of the linac's beams at high energies are not sensitive to longitudinal 
space charge effects for existing beam currents. The crystalline or 
ordered beam have of course many interesting and important properties 
and applications. A few properties and one very important application is 
considered in this paper. It is clear that particles of dense bunches may 
be arranged in the certain orders. In these ordered bunches particles 
form transverse planes, which are same spacing in the longitudinal 
direction. The conditions and properties of such ordering are considered 
in the next section of this paper. Such bunches are radiate coherently at 
the wavelengths integer times smaller than interplane distances, as shown 
in section 3 of this report. The importance of tunable and powerful sources 
of the coherent radiation of XUV wavelengths has stipulated several 
projects of FEL in recent years \cite{1,2}. Those FEL's operate in the Self 
Amplified Spontaneous Emission (SASE) mode, i.e. starting from noise in 
the initial distribution of the electron beam longitudinal density. This 
evolution determines the undulator length needed to reach saturation 
practically to be very long (about hundred meters). The necessary 
power of laser radiation is possible to obtain by using ordered 
bunches. This radiation calls superradiant regime of FEL radiation 
\cite{3-7} or coherent spontaneous emission (CSE). As shown in above 
references it is possible in two cases. First, for short bunches and 
second for modulated bunches. The superradiant regime due to self 
bunching of the beam in the FEL is considered by Bonifacio \cite{6,7}. 
In difference from above references where is considered the coherency 
of long wavelength radiation (the wavelength much more than mean 
distance between bunch electrons), in this report is considered the 
coherent radiation of the relativistic bunch when the radiation wavelength is 
about or less than distance between particles. In this case it is important 
the discreteness of beam and correlations between beam particles 
positions, because of strong Coulomb interaction of the beam particles.

\section{Correlations and Ordering of the Relativistic Bunches}

\subsection{Introduction} The requirement of bunch uniformity  assumes, 
that electrons of the bunch with density n are placed on the same mean 
distance $a=n^{-1/3}$ from each other. Such a placement is possible only when 
each three of particles compose equilateral triangle as a result of strong 
correlation between particles positions. This is connected with approximately 
hexagonal structure of the disordered medium \cite{8} as the most probable.
Each particle in a such medium have 14 nearest neighbors, but numerical 
calculations show that the mean number of nearest neighbors is about 15 with 
mean deviation about 10 percents. For medium with the long distance 
interparticle interaction such as beam or bunch the vertices of triangles are 
equilibrium points of the particle positions. Particles which are not 
placed in their equilibrium points oscillate as shown below around their 
equilibrium positions with amplitudes equal to deviations from equilibrium 
points. It is well known that properties of any medium depend on Madelung energy -  
dimensionless parameter $\Gamma$, which is the ratio of the depth of the 
interparticle potential well and medium temperature. For neutral and one 
component plasma (OCP) due to the Debay screening interparticle potential 
is about bare two particle Coulomb potential $e^2/a$ \cite{13}. The numerical 
Molecular Dynamics (MD) calculations show that for $\Gamma\geq 172$ 
 OCP is crystallize with body centered cube (bcc) lattice \cite{12}.
Detailed MD calculations \cite{11} for finite full charged Coulomb 
systems with number of particles about thousand was show that in this 
case crystallization takes place at low values of $\Gamma$, but this 
result has not been explained theoretically yet. We show in this report that for 
Coulomb systems of high energy charges the potential well of each particle 
is for about $N_t$ (number of particles in the one transverse plane of the 
bunch) times deeper than that of the bare two particle Coulomb potential.    
In contrast to crystalline beams in the storage rings, where strong 
cooling is necessary for attaining crystalline state \cite{15}, as shown 
in the \cite{18} a such beam optics may be designed that energy spread,
particle dynamics and sizes of the  bunch in the Final Focus of the linac
are sensitive to longitudinal space charge effects only for bunches with 
$N\succeq 10^{11}$. Calculations show that the beam of the SLAC in 
the Final Focus have $\Gamma$ for about a few hundreds that means 
that bunch may be crystallized. Note, that in this case resonances with 
betatron oscillations are not important because of longitudinal ordering 
of the bunch and disordering in the transverse directions.
Classical consideration is valid, when $\hbar\Omega\ll T=mc^2 \delta^2$,
with $\delta=\frac{\Delta E}{E}$ relative energy spread.This condition is equal 
to $\varepsilon=\frac{\lambda _c \sqrt{\Gamma}}{a\delta\gamma^{3/2}}\ll 1$,
which is satisfied for $\gamma\sim 10^4$. In this case bunch become 
longitudinal ordered as considered in the section 2.3. When the condition 
$\varepsilon\leq 1$ satisfied the quantum effects should be taken in the 
account and results of the \cite{16,17} can be used for bunch electrons 
correlation function. This case is considered in the section 2.2.

\subsection{Correlations of the Strong Interacting Charges Ensembles}
  
  The statistical properties of N particle ensemble are described by 
N-particle distribution function $f_N$, which is depend  generally
 on positions and momentums of all N particles of the ensemble. As 
seen below for calculating of the radiation coherency the two particle 
distribution function 
$$f_2=\int d^{N-2}xd^{N-2}pf_N$$,
or the correlation function is necessary
$$f_2 =g_2 (x_1,x_2)+f_1(x_1)f_1(x_2)$$.
 Charges in the bunches are strong correlated because of strong Coulomb 
interaction. As considered above, quantum effects should be taken into account,
when $\varepsilon$ do not much less than unity.
In this case we can use the density correlation functions obtained in the 
\cite{16,17} for strong correlated electrons
\begin{equation}
\langle \varrho(x)\varrho(0)\rangle \approx cos(2\pi \frac{x}{a})exp[-b\sqrt{ln(x)}]
\label{corr}
\end{equation}
where $b=\frac{\varepsilon}{\Gamma}$, m is effective mass,in the units
$\hbar=1$.It is clear that  $b\ll 1$ and one can obtain from 
this correlation function the approximate two particle distribution 
function
\begin{equation}
g_2(x_1,x_2)=(1+\cos{2\pi\frac{x}{a}})/l
\label{ordfunc}
\end{equation}
where $l$ is the bunch length.This is true in the liquid or partially ordered 
state of the bunch. Such state is possible to obtain at energies of the electrons 
for about tens Mev, if the bunch have energy spread for about percents and 
number of particles about $10^{10}$, which is about ten times greater than 
that in SUNSHINE \cite{19}

 \subsection{Ordering of the Charges in the Relativistic Bunch}
 After expanding of the full force between particles with charge e and
fixed distance R moving along longitudinal z direction with velocity 
$v=\beta c$ and angle $\theta$ between R and z directions,
\begin{eqnarray}
F_z=\frac{e^2}{R^2}\frac{(1-\beta^2)cos{\theta}}{(1-\beta^{2}(sin{\theta})^2)^3/2}\\
F_\perp=\frac{e^2}{R^2}\frac{(1-\beta^2)sin{\theta}}{(1-\beta^{2}sin^2{\theta})^{3/2}}
\end{eqnarray}
around equilibrium points of the bcc lattice,one can find that potential wells in 
the longitudinal and transverse directions are 
\begin{eqnarray}
U_{0\|}=U_{0}\frac{N_t}{4\gamma^2}S_l&,&S_l=\sum{\frac{1}{n^3}}\\
U_\bot=U_0\frac{a_l}{a_t}\frac{\gamma}{4} S_t &,&S_t=\sum{\frac{n_1}{(n_1^2+n_2^2)^2}}
\label{pot}
\end{eqnarray}
  
where $U_0=e^2/a$ is the bare Coulomb potential,$N_t$ is the number of 
particles on each transverse plane.The sum in the $S_l$ is carrying out over 
number of transverse planes ($n_3 $ denotes the plane number),and the sum 
in the $S_t$ over $n_1,n_2$ which numerate particles in x, y directions 
respectively on the each plane.Here is assumed that self space charge forces 
of the bunch are compensated by external restoring forces \cite{11,14}. In this 
case particles oscillate in the potential 
\begin{equation}
U_z =U_{0\|}\frac{z^2}{a^2}
\label{pot}
\end{equation}
Such deepening 
of the potential well in the bunches in comparison with known potential 
well in the OCP  means that crystallization or ordering of bunch is possible 
at comparable higher temperatures than of OCP. This we proof by consideration 
of the canonical partition function of the Gibb's canonical ensemble with 
hamiltonian $H_N$ and inverse temperature in energy units $\beta=1/kT$: 
\begin{equation}
Q_N(\Omega,\beta)=\frac{1}{h^{3N}N!}\int dr^Ndp^Nexp{(-\beta H_N(r^N,p^N;\Omega))}
\end{equation}
for a system of N charges enclosed in a volume $\Omega$; here h is Planck's 
constant. In the approach used by us hamiltonian $H_N$ of N relativistic charges 
is represented in the usual classical form as a sum of kynetic $E_N$ and 
potential $U_N$ energies $H_N=E_N+U_N$. Total potential energy $U_N$ 
equal to the sum of  one particle potentials \ref{pot}. In this approach the 
usual Helmholtz free energy $F_N(\Omega, \beta)$ as thermodynamic potential 
\begin{equation}
Q_N(\Omega,\beta)=exp{(-\beta F_N(\Omega, \beta))}
\end{equation}
and dimensionless configuration integral 
\begin{equation}
Z_N(\Omega,\beta)=\int \frac{dr^N}{\Omega^N}exp{(-\beta U_N(r^N,\Omega, ))}
\end{equation}
It is easily seen from this equation that $Z_N$ depends on $\beta$ and on 
interactions only through the dimensionless combination $U_{0\|}\beta$, which 
is the usual OCP $\Gamma$ multiplied by $N_t/\gamma^2$. So, the $\Gamma$ 
for dense enough ensemble of relativistic charges (bunch), can be greater than 
$\Gamma$ for OCP. 
Particle of the bunch in this potential well is oscillate with frequency 

\begin{equation}
\Omega ^2 = \frac{U}{2ma^2}
\end{equation}
where a is the mean interparticle distance in correspondence direction,
m is the effective mass of bunch particle equal to $m\gamma^3$ for
linacs . This formula is obtained from equation of particle motion in the 
longitudinal potential \ref{pot}
\begin{equation}
\label{eqmotion}
m\ddot{z}+\frac{U_{0\|}}{a^2}z=0
\end{equation}

The case of real storage ring lattice was considered in \cite {15}. If the 
crystallization time which equal few oscillation periods less than period 
of betatron oscillation of the accelerator the bunch becomes ordered. In the 
harmonic approach the potential in the longitudinal direction is equal
\begin{equation}
U_\|=U_{0\|}\frac{z^2}{2a^2}
\end{equation}
This formula show that the decreasing of the $a$ means increasing of the 
potential well. Such decreasing of the interparticle distance take place in the 
linacs due to adiabatic decreasing of the bunch sizes or bunch compression.
Notice, that since magnetic compression do not change full energy, the bunch 
temperature (mean kinetic energy in the rest frame) must be decreases 
due to ensemble full potential energy (which is positive for the ensemble of 
one sign charges) increasing because of compression. More detailed 
analyze show that during acceleration in the linac $\Gamma$  is changes 
in correspondence with dependence 
$$\Gamma\sim \gamma ^{1/2}l^{1/2}$$
Such result is obtained from the assumption of adiabatic changing of the 
energy $\gamma$ and bunch longitudinal size l. This is possible in the 
case large $\Omega$, which may be obtained after sufficient compression 
of the bunch.

\section{Coherence Radiation of the Bunch}

\subsection{Introduction}
Radiation of the ensemble of N  electrons moving along identical trajectories 
but with arbitrary spatial placement may be written in the form \cite{3}
\begin{equation}
I=I_0 N(1+(N-1)F)
\end{equation}
where $I_0$ is the intensity of single electron radiation,N is the number of the  
electrons in the bunch,F is the factor of coherency of the bunch and may 
be written in the follow form \cite{3}
\begin{equation}
F=\frac{1}{N^2}\sum{e^{i\vec{k}\vec{r_j}}}\sum{e^{-i\vec{k}\vec{r_j}}}
\label{ff}
\end{equation}
The position of the j-th electron $\vec{r_j}$ can be written as a
sum of transverse and longitudinal parts $\vec{r_j}=\vec{r_{jtr}}+z_j$.
For longitudinal waves the transverse part of the F is equal to $N_r^2$.
This is right because of transverse coherency of the radiation \cite{3}. Note,
that generally this expression must be averaged over all possible ensembles,
or over N-particle distribution function. But as seen from above formulae for 
coherency formfactor this one depend on two particle coordinates, so we can 
use for averaging the two particle distribution function. After such calculations 
formfactor in our notation is equal to 
$$F=F_0 +F_1,$$
where $F_0$ is the usual formfactor \cite{3} and $F_1$ is the additional term 
due of averaging over correlation function g. This term describe the effect of 
correlations on the bunch coherency. For known correlation function formfactor 
of correlations can be calculated by averaging of the general formulae by 
correlation function.

\subsection{Coherent Radiation of the Strong Correlated Uniform Bunch}
 
  When the condition of crystallization or full ordering of the bunch 
is not satisfies,but the interparticle interaction is strong enough,
the bunch becomes ordered with correlation function (\ref{corr}). In
more simple notation we can write for factor b
$$b=\frac{\lambda _c}{a\delta \sqrt{\Gamma(\gamma)^3}}$$ 
where $\lambda _c $-Compton wavelength,$\delta =\frac{\Delta E}{E}$ -mean 
energy deviation. It is not difficult to see, that for physical interesting
range of parameters b is less than $\lambda _c/a$, and much less than unity.
Consequently, we can the neglect exponential dependence in this formula and 
use (\ref{ordfunc}) for two particle distribution function.
 In this approach the longitudinal coherency formfactor is equal
\begin{equation}
F=\frac{N_r^2}{N^2}\frac{\sin^2{Nx}}{x^2}
\end{equation}

\subsection{Coherent Radiation of the Ordered Bunch}
  Consider now radiation of the ordered bunch. For $N_z$ planes with $N_r$ 
electrons on each plane the coherency factor is may be written in the following 
form
$$F=\frac{N_{r}^2}{N}\sum{e^{ikj\tilde{a}}}\sum{e^{-ikj\tilde{a}}}$$
where  $\sum{}$ -means sum over planes in longitudinal direction.
It is not difficult to obtain that
$$F=\frac{N_{r}^2}{N}\frac{\sin^2{N_{z}x}}{\sin^2{x}} $$
and when the value $x=ka$ is approach to $n\pi$ (resonance condition) the 
coherency factor or laser gain becomes about $N$. This consideration is 
true for bunch with step function for longitudinal density distribution 
(for homogeneous or uniform bunch)
  
\subsection{The Influence of fluctuations}

Now consider the influence of fluctuations or deviations of the 
particle positions from hexagonal model above on the coherence short
wave length radiation of the bunch. Let us assume that particle is
on the mean distance, but not in the z-direction. In this case the
particle position is fluctuated on the value $a \theta^2/2$, and
if $x\theta^2/4\ll 1$ we may neglect this fluctuation. Notice,
that in important practical cases this condition is satisfied. It
is not satisfied only for very short radiation when $x\gg 1$.
The second source of fluctuations from our model is possible, when
the particle is replaced on the z-direction but not in the
distance a. In this case F multiplied by the factor
$e^{-k^{2}b^2}$, where b is the dispersion of the fluctuation. It is
clear that if $b/\lambda\ll 1$, this fluctuations not disturb the
radiation coherency.Executed numerical calculations show that 
fluctuations about ten times smaller than inter particle distance $a$.

\section{Conclusions}
So, in this paper the influence of bunch particle position correlations on the 
bunch undulator radiation are investigated. Additional term in the radiation 
coherency formfactor was found. The correlations in the relativistic charge 
bunches are considered and are found the crystallization or ordering conditions
of dense relativistic bunches or beams, which may be satisfied much easier 
than those for OCP. This very exotic and interesting state of  matter can be 
obtained at existing linac beams. We show that ordered bunches can radiate 
coherently ,i.e. much powerful than spontaneous one. In dependence 
of beam energy and density this radiation may have wavelength at XUV
range also.

\end{document}